\documentclass[prd,preprintnumbers,twocolumn,amsmath,amssymb,nofootinbib,showpacs]{revtex4}

\usepackage{epsfig,amssymb,amsfonts,verbatim}

\def\bfl{\begin{flushleft}}
\def\efl{\end{flushleft}}
\def\bfr{\begin{flushright}}
\def\efr{\end{flushright}}
\def\bc{\begin{center}}
\def\ec{\end{center}}
\def\be{\begin{equation}}
\def\ee{\end{equation}}
\def\ba{\begin{eqnarray}}
\def\ea{\end{eqnarray}}
\def\baa#1{\begin{array}{#1}}
\def\eaa{\end{array}}
\def\bw{\begin{widetext}}
\def\ew{\end{widetext}}
\def\nn{\nonumber }
\def\lb#1{\label{#1}}

\def\drm{\text{d}}

\def\schrod{Schr\"odinger  }
\def\reno{Reissner-N\"ordstrom  }

\newcommand{\scopr}{SCP}

\def\lan{{\cal L}}

\def\ap{{A}}

\newtheorem{Def}{Def.}[section]

\newtheorem{Prp}[Def]{Proposition}

\def\k{\kappa}

\def\m{\mu}

\def\n{\nu}

\def\s{\sigma}

\begin{document}

\preprint{hep-th/0512128}

\title{
Generic approach to 
dimensional reduction and 
selection principle for low-energy limit of M theory
}

\author{Konstantin G. Zloshchastiev}

\affiliation{Instituto de Ciencias Nucleares, 
Universidad Nacional Aut\'onoma de M\'exico, A.P. 70-543,
 M\'exico D.F. 04510, M\'exico}
 
\affiliation{Department of Physics, National University of Singapore,
Singapore 117542 
}
 



\begin{abstract}
We propose the approach to deriving lower-dimensional limit of modern 
high-energy theory which does not make explicit use of 
the Kaluza-Klein scheme and predefined compactification manifolds. 
The approach is based on the selection principle in which a crucial role is played by p-brane solutions
and their preservation, in a certain sense, under dimensional reduction. 
Then we engage a previously developed  method of reconstruction of a theory 
from a given solution which eventually leads to some model acting in the space of field couplings.
Thus, our approach focuses on those general features of effective 4D theories 
which are independent of how the decomposition of spacetime dimensions into ``observable''
and ``unobservable'' ones could be done.
As an example, we exactly derive the simplified abelian sector of 
the effective low-energy M-theory together with its fundamental 0-brane solution describing the family of charged 
black holes with scalar hair in  asymptotically flat, de Sitter or anti-de Sitter spacetime.
\end{abstract}

\maketitle



\section{Introduction}

Nowadays it is  believed that the most promising candidate for a unified theory
of interactions
is no longer ten-dimensional (10D) string theory 
but rather 11D M-theory.
Its explicit formulation is still pending but one knows some low-energy limits,
the 11D supergravity (SUGRA) and 10D superstring theories.
Yet, 
the M-theory 
remains so far
an abstract theory
whose predictions can not be 
experimentally tested in  foreseen future.
The main reason is that one does not know yet how the low-energy
4D limit of M-theory should look like:
in the standard approach to dimensional reduction, 
the Kaluza-Klein (KK) compactification scheme,
the ambiguity arises
due to huge variety of possible manifolds to compactify on 
to get down to 4D \cite{Polchinski:1998rr}.
This ambiguity is not surprising though - in its recent form, the KK approach is more a mathematical procedure 
than a physical law and thus  need not to be unique. 

Among other issues, the KK ambiguity leads to the well-known ``cosmological constant problem'' -
why this important parameter has a non-zero value
(despite being exactly zero on the supermoduli space), and how to determine
this value.
In string/M-theory the problem arises due to the non-uniqueness 
of the corresponding 4-form field strengthes \cite{Brown:1988kg}:
consider a compactification of M-theory from
11D
to 4D using, e.g., a
7-torus. The torus has a number of moduli
representing the sizes and angles between seven 1-cycles.
The 4-form fields have as their origin a fundamental 7-form field strength
of M-theory. These 7-forms can
be chosen such that three of their indices are identified with
compact dimensions, and this 
can be done in  $7 !/(3 ! \, 4 !)=35$
ways  which produce that many distinct
4-form fields in the uncompactified space.
More generally, in the kinds of compact manifolds
used to reproduce the Standard Model 
there can be many independent ways of wrapping three compact
directions with flux and thus producing a great number of four-dimensional 4-form fields.



To illustrate how all aforesaid affects four-dimensional physics, let us consider 
gravity $g_{\mu\nu}$ coupled to 
electromagnetic ${\cal A}_\mu$ and scalar $\phi$ fields given by the action (in the Einstein frame):
\be
 S = \int \drm^4 x \sqrt{- g} \, 
\biggl[
      R - \frac{1}{2} \,(\partial \phi)^2 +
      \Xi (\phi) \,F^2 + \Lambda (\phi)
\biggr],                                                        
\lb{eEMD}
\ee
where
$F_{\mu\nu}= \nabla_\mu {\cal A}_\nu - \nabla_\nu {\cal A}_\mu$;
we use the units where $16 \pi G = c = 1$, where $G$ is the
Newton gravitational constant.
The particle content of this theory (graviton, photon and scalar) is
capable of describing the bosonic part of 
many physical phenomena, from electromagnetism to cosmology.
It is thus not surprising that (\ref{eEMD}) appears in the abelian sector of the
effective 4D low-energy limit of M-theory (EFMT).
The open question here is what are  explicit values of 
the Maxwell-dilaton coupling $\Xi$ 
and the scalar field self-coupling $\Lambda$ \cite{Damour:1994zq}.
In the tree-level approximation (spherical topology
of intermediate worldsheets)
string theory suggests that  
\be                                                 
\Xi^{(0)} = - e^{- a \phi}, \ \
\Lambda^{(0)} = 0,
\lb{e-minstr}
\ee
at the leading order in
the inverse string tension \cite{Fradkin:1984pq,Callan:1986jb}.
Obviously, a model with such couplings is not phenomenologically satisfactory - 
one can not describe, for instance, cosmological phenomena.
The realistic $\Xi (\phi)$ and $\Lambda (\phi)$ 
are different from these for at least
two reasons.
First, the 
EFMT being an effective 4D theory 
is supposed to hold information
about higher dimensions and fields, therefore,
in the case of the model (\ref{eEMD})
this information must be stored in 
$\Xi$ and $\Lambda$ thus affecting their form.
Second, 
the couplings will get modified also by quantum corrections to
the initial 10, 11D Lagrangians, 
see, for example, Refs. \cite{Dixon:1991pc,Cvetic:1993ty}.
In short, to construct the genuine EFMT, ideally one  should perform two steps: 
(i) construct the ultimate  M-theory, 
and then (ii) dimensionally reduce it down to 4D.
Both steps are obviously of enormous difficulty.
 
On the other hand,  
vast experimental data 
gave so far no evidence for higher-dimensional physics  
to a high degree of precision.
It is thus a good question how the Nature ``chooses'' that very unique way 
of dimensional reduction which determines our essentially 4D Universe.
Is there any selection principle (apart from anthropic arguments)
by which one could make the unique choice?

\section{Branes and selection principle}\lb{sec:scp}

Here we try to formulate one. 
Suppose, an   
observer wants to describe a certain physical phenomenon or 
object (black hole, for example). 
According to an M-theorist the complete description of the 
object is given by a high-D solution, call it  
$Sol(M)$, of a certain high-D theory $M$. 
However,  
our observer can experimentally operate only in four dimensions -
his apparatus (including organs of sense) always gives essentially (3+1)-dimensional output, hence, 
he never perceives extra dimensions directly
but only as additional ``forces'' in the 4D framework.
The main aim of EFMT thus is to incorporate these corrections while staying compatible 
with the four-dimensionality of experimental data. 

In other words, one 
needs the dimensionally reduced description of the object, i.e., 
the one in terms of the most appropriate solution $Sol(M_4)$ arising from a 
corresponding effective 4D theory $M_4$. 
The question is how to derive such EFMT $M_4$, especially assuming the situation like
the one we have recently, i.e., when $M$ is not explicitly known?
Our solution to the problem would be to use the known fundamental 
solutions of high-D theories as a guiding thread.
Indeed, real phenomena are described by $Sol(M)$'s thus if one wishes to preserve
(partially, at least) such 
description in the dimensionally reduced theory one must make sure 
that the latter  does not disallow $Sol(M)$-like solutions in principle.

Considering aforesaid, 
we formulate the following selection criterion:
``Suppose 
we have a higher-dimensional theory, call it $M$.
This theory has a certain  
physically relevant 
solution, $Sol(M)$ which is unique to $M$.
This solution has certain distinctive property, $Prop(Sol(M))$, 
which 
is preserved under reduction of dimensions.
On the other hand,
we have 4D theory $M_4$ whose nontrivial solutions we denote by $Sol(M_4)$.
Then the necessary condition for $M_4$ to be a lower-dimensional limit of $M$
is that at least one of $Sol(M_4)$ must have  $Prop(Sol(M))$.'' 
This criterion 
can be regarded as
the correspondence principle for solutions, henceforth we call it the 
\textit{Solution Correspondence Principle} (\scopr).
Now, 
to determine
what solution is physically relevant and 
which property of it is suitable for \scopr,
we will make use of the following two facts from the string/M-theory.

First, 
it is well-known that 
the branes are 
known
solutions of 
M theory (one may recall M2 and M5 ones) hence they should be perfectly suitable
for a concrete realization of \scopr.
Moreover, it is known that branes are inevitable for proper describing of 
black holes (microstates, entropy, etc.) \cite{Strominger:1996sh}, therefore, their absence in a theory
would cause serious difficulties with consistent explaining such issues.
In turn, absence of black holes would lead to the loss of protection of a theory from ubiquitous
appearances of naked 
singularities \cite{Penrose:1964wq} and thus to undesirable violations of the
Cosmic Censorship principle \cite{Penrose:1969pc,Wald:1997wa,Brady:1998au}. 
In what follows
we concentrate solely on the most primitive type of branes, 
the p-branes, as they are minimal-energy configurations and  can be viewed as a kind 
of ``attractor points'' in the evolution
space of all brane solutions. 
The metric of a generic  p-brane 
can be schematically described as
\be\lb{eBsch}
\textbf{g}_D = 
e^A \, \breve{\textbf{g}}_{\tilde d} + \text{ the rest}
,
\ee
where 
$\breve{\textbf{g}}_{\tilde d}$ is the metric of the ($D-p-2$)-dimensional 
transverse 
space,
$e^A \equiv R_{\text{trans}}^2$ is the warp factor or effective squared radius of transverse dimensions;
``the rest'' is usually the ($p+1$)-dimensional Poincar\'e-invariant
metric plus $g_{rr} (r) \drm r^2$ with $r$ being the isotropic radial coordinate
in the transverse space.
The brane solution also has a scalar field (dilaton) part which is represented simply
by $\phi=\phi(r)$, and a gauge field part which is of no interest to us here.

Second, we know that the dilaton effectively represents
the dynamically varying string coupling constant; from the viewpoint of 
the 11D M-theory it is related to the size of the 11th dimension.
Note that the dilaton is defined as the scalar non-minimally coupled to gravity.
Then if one assumes that the spacetime is flat (as in the Standard Model),
one can not have the non-minimal term in the SM Lagrangian anyway.
What is left is the kinetic and potential terms for the ``dilaton'', therefore,
the dilaton, even if it existed in 10D supergravity, in the flat-spacetime limit becomes a 
normal (minimally coupled) scalar field
(e.g., a long-range background field such as the inflaton).
Moreover, in the low-energy limit, even if spacetime curvature does not vanish, 
one can always conformally rescale the metric 
such that inside the action any $\phi$-dependent factor in front of 
the Ricci scalar  can be absorbed into metric.
Of course, it changes the frame we are working in from the string to the Einstein one 
but here it is not so important as long as phenomenological matter is not involved.

Thus, we have two fundamental quantities: the one of physical origin, $\phi$,
and the one related to multi-dimensional geometry, $e^A$. 
Consequently, the relation $\phi=\phi(R_{\text{trans}})=\phi(e^A)$ is a scalar one 
and contains certain 
information about the theory, for example, about
how branes' properties would change 
at different size of transverse dimensions.
Thus, such relation must be a universal property of brane solutions -
in particular, 
it is unlikely that the reducing of dimensions will drastically change it.
Therefore, we expect that the function $\phi(e^A)$, 
or, alternatively, 
the inverse 
$A(\phi)$, 
is \textit{structurally invariant}, 
i.e., has the same functional dependence 
(except perhaps for different constant parameters)
for any fundamental p-brane 
and at any physically admissible $D$.
Matching notations with \scopr, we associate branes 
with $Sol(M)$ and $A(\phi)$ with $Prop(Sol(M))$, thus one can restate \scopr \,  
as: ``\textit{the genuine EFMT should necessarily
possess at least one solution of the p-brane type}''. 

To check the structural invariance of $A(\phi)$ and at same time to derive its explicit 
form, consider the typical theory giving rise
to brane solutions, truncated supergravity.
Its 
abelian sector is described by the Lagrangian 
\be\lb{st-lagr}
\lan \sim R - \frac{1}{2} (\partial \phi)^2 - (1/2n !)e^{a \phi} F^2_{[n]},
\ee
where $F_{[n]}$ is the antisymmetric tensor of rank $n$
\cite{Lu:1995yn,Duff:1996hp,Duff:1996qp}.
The corresponding p-brane solution 
respects the 
$\hbox{
(Poincar\'e)}_d\times {\rm SO}(D-d)$ 
symmetry and is given by (omitting the gauge field part):
\be
ds^2 = H^{-4\tilde d\over\Delta(D-2)}dx^\mu dx^\nu\eta_{\mu\nu} +
H^{4d\over\Delta(D-2)}dy^mdy^m, \lb{st-pbrme}
\ee
\ba
e^\phi &=&
H^{2a\over\varsigma\Delta},\ \
H \equiv 1+{\k\over r^{\tilde d}},
 \ \
~^{\mu = 0,1,..,p,}_{ m=p+1,..,D-1} \ ,
\lb{st-pbrphi}
\ea
where 
$r=\sqrt{y^my^m}$,
$
a^2 = \Delta-{2d\tilde d\over D-2}, \
d=p+1,\
\tilde d = D-d-2,
$
$\varsigma = \pm 1$ for the electric and magnetic branes, respectively;
$\k$ is the integration constant which sets the brane's mass scale 
(its positiveness ensures the absence of naked singularities at finite $r$).
Using ${\rm SO}(D-d)$ symmetry, one can rewrite the r.h.s. of Eq. (\ref{st-pbrme})
as
$
H^{-4\tilde d\over\Delta(D-2)}dx^\mu dx^\nu\eta_{\mu\nu} +
H^{4d\over\Delta(D-2)} dr^2 + r^2 H^{4d\over\Delta(D-2)} d\Omega_{(\tilde d-1)}^2.
$
It is the last term we are interested in -
comparing it to the first term on r.h.s. of Eq. (\ref{eBsch})
we can identify
$ 
e^{A} $
with
$ r^2 H^{4d\over\Delta(D-2)}.
$
Eliminating $r$ using Eq. (\ref{st-pbrphi}) we finally obtain
\be
A (\phi) =
\frac{2\,d\,\varsigma }
   {a\left(  D -2 \right) } \phi
+
  \frac{2}{{\tilde d}} \ln \k
-
  \frac{2}
     {{\tilde d}}
     \ln (
       e^{\frac{\Delta \,\varsigma \,\phi }{2\,a}} -1)
,
\lb{eBC}
\ee
i.e., $A(\phi)$ has the form $c_1 \phi + c_2 + c_3 \ln{(\text{e}^{c_4 \phi}-1)}$,
where $c_i$ are certain $D$-dependent constants, and one can check that such relation holds for
any physically admissible (such as those free of naked singularities, etc.) p-brane solution known so far.
Thus, $A(\phi)$ has the same structure for any $D$, the only things to change
are the values of $c_i$'s.  

Now, according to \scopr $\,$ the necessary condition for the theory to be a genuine 
EFMT is that at least one of its  solutions must have the property (\ref{eBC}).
If as a skeleton model we take Eq. (\ref{eEMD}) then
our task now is to determine  unknown couplings 
$\Xi(\phi)$ and $\Lambda(\phi)$.
Therefore, we demand that its 
simplest fundamental (static and spherically symmetric) 
solutions have the property (\ref{eBC}) evaluated at $D=4$.
In 4D the only choice is $p=0$, $d=\tilde d =1$ (0-brane)
so Eq. (\ref{eBC}) takes the form
\be
A (\phi)_{D=4} =
(a \varsigma)^{-1} \phi
+  2 \ln \k
- 2  \ln ( e^{\frac{a^2-1}{2 a \varsigma}\phi} -1)
.
\lb{eBC4}
\ee
The next step thus would be to find a class of 4D theories
whose solutions 
resemble 
the higher-dimensional p-branes
in a sense of having the property expressed by the last formula.
One should not confuse this with demanding that the whole higher-dimensional brane's
structure (isometries, etc.) should be preserved in 4D - it is unlikely possible
in general case.
Fortunately, from the viewpoint of SCP such demand would be too strong and redundant (notice
that SCP has been defined as a necessary but not sufficient condition).
Instead, we accentuate on deriving an effective 4D theory whose solutions 
mimic the brane in just one property (\ref{eBC4}) but the theory must be as general as possible
(within the initial restrictions made upon its field content, of course).
The existence of at least one such  solution will be proven
subsequently by construction -
in fact, in 4D the spherical symmetry and existence of time-like Killing vector
guarantee the appearance of the solutions of the form (\ref{eBsch}) as 
the latter becomes just the (isotropic) gauge condition for spacetime metric.
Incidentally, because of the latter circumstance, $D=4$ becomes a distinguished number in this sense. 

\section{Brane class and couplings}

Now,
how to reconstruct the theory whose 
solution (one, at least)
would possess the property given by Eq. (\ref{eBC4})?
The related question is how large would  the corresponding
class of equivalence be, i.e., how many theories
exist which have at least one such solution
(this is especially crucial for  uniqueness of the EFMT
defined in such a way)?

More technically: we have the theory defined by Eq. (\ref{st-lagr}) for arbitrary $D > 2$
which has the generic p-brane solutions with the property (\ref{eBC}).
However, this property can be more ``generic'' than the theory itself, in a sense 
that there may exist Lagrangians more general than (\ref{st-lagr}) such that their 
appropriate solutions (one, at least) still obey Eq. (\ref{eBC}).
How to reconstruct the bijective preimage (kernel of the homomorphism), i.e., 
the \textit{most general}  form of 
such Lagrangians assuming $D=4$ and restricting oneself to
a particular skeleton action, e.g., Eq. (\ref{eEMD})
with the couplings $\Xi$ and $\Lambda$ to be determined?

Luckily, we have  quite general theory addressing all these issues. 
In Ref. \cite{Zloshchastiev:2001da} it has been developed 
for the abelian theory (\ref{eEMD}) but definitely the idea should work in more general case as well.
It surprisingly turns out that the ambiguity is not big at all:
the unknowns $\Xi(\phi)$ and $\Lambda(\phi)$ can be determined uniquely
up to two constants each (apart from the freedom to select a frame by conformal rescaling of  metric).
To see this, consider the field equations following from action (\ref{eEMD}) 
in the static spherically symmetric limit (where the metric takes the form (\ref{eBsch})).
Using them, one can derive 
the following equation \cite{Zloshchastiev:2001da}:
\ba
&&
\frac{\Upsilon '}{\ap '}
+ \frac{1}{2}
\left(
\frac{1}{ \ap '^2} + 1
\right)
\Upsilon
+
\frac{e^\ap}{2}
\left(
      \Lambda + e^{-2 \ap} \hat\Xi
\right)
=- 1,                         \lb{e-g1a}\\&&
\Upsilon \equiv
\frac{
1   +
e^\ap
(
      \Lambda +  \Lambda ' \ap '
)/2
+
e^{- \ap}
(
      \hat\Xi +  \hat\Xi ' \ap '
)/2
}{
(1/\ap ')'}
,\nn\\&&
\hat\Xi \equiv 2 \, (Q^2\,\Xi^{-1} + P^2 \, \Xi ),
\nn
\ea
where
constants $Q$ and $P$ stand for electric and magnetic charges, 
respectively; throughout the section prime stands for derivative 
with respect to $\phi$.
Now, suppose we have fixed  $\ap (\phi)$,
then the equation turns into a joint linear second-order ODE
with unknown functions $\Lambda (\phi)$ and $\hat\Xi (\phi)$.
This ODE (called the \textit{class equation}) is, in fact,
the constraint for $\Lambda$ and $\Xi$ which ensures 
the internal consistency of the theory.
Therefore, with every given $\ap (\phi)$  
one can associate the appropriate 
class of solvability 
given by a self-consistent 
$\{\Lambda,\,\Xi\}$ pair.
Thus, the space of all possible coupling functions becomes 
``inhomogeneous'' as it
can be divided according to the class structure.
Henceforth we call the class defined by Eq. (\ref{eBC})  the 
\textit{p-brane class of equivalence}.
Theories belonging to this class are equivalent in a sense
that their appropriate solutions possess the same dependence $\ap (\phi)$.
For $D=4$, substituting Eq. (\ref{eBC4}) into Eq. (\ref{e-g1a}), 
we obtain:
\ba
&&
\hat\Xi ''
+ \varsigma
\left(
a - \frac{1}{a}
\right)
\hat\Xi '
-
\hat\Xi
+
      \left(
\frac{\kappa
      e^{\frac{\phi}{2 a \varsigma}}
     }
     {
            e^{\frac{(1+a^2) \phi}{2 a}} - 1
     }
      \right)^4
\times \qquad
\nn\\&&
\biggl\{
\Lambda ''
-
\left(
a + \frac{1}{a}
\right)
\text{coth}\!\left[
\frac{(a^2+1)\phi}{4 a}
\right]
\Lambda '
+
\Lambda
\biggr\}
=0 
.  
\lb{joinedce}                                                   
\ea
It contains $\k$ which came from Eqs. (\ref{st-pbrme}), (\ref{st-pbrphi})
where it was a mass scale parameter.
Such constants are  attributes of solutions 
and thus should not explicitly appear in the action,
in the same manner as, for example,
mass is not built into the couplings in a field Lagrangian  but
appears only inside, say, the \reno solution, i.e., as an integration constant.
The only way to get rid of $\k$ is to separate the equations:
\ba
&&
\hat\Xi ''
+ \varsigma
\left(
a - \frac{1}{a}
\right)
\hat\Xi '
-
\hat\Xi
=0,
\lb{eCE1}
\\&&
\Lambda ''
-
\left(
a + \frac{1}{a}
\right)
\text{coth}\!\left[
\frac{(a^2+1)\phi}{4 a}
\right]
\Lambda '
+
\Lambda
=0 
,  
\lb{eCE2}                                                   
\ea
i.e., the physical condition of independence of the skeleton
action from the mass scale parameter $\k$ naturally implies that the total coupling space 
(defined ``on top'' of  the space of fields) 
factorizes into the direct product of ``elementary'' coupling spaces:
$\{ \Lambda,\, \hat\Xi \} \to \{ \Lambda \} \otimes \{ \hat\Xi \}$.
The coupling functions can be easily found by solving these ODE's.
For arbitrary $a$ we have:
\ba
&&
\hat\Xi (\phi) = 
\frac{1}{2}
\left(
\s_1 e^{\phi/a} + \s_2 e^{-a \phi}
\right)
,
\label{eXi-G}\\&&
\Lambda (\phi) = 
\sinh^2 (\phi/\phi_1)
\bigl[
C_1 P_2^\mu (z)
+
C_2 Q_2^\mu (z)
\bigr]
,
\label{eV-G}
\ea
where $P_\n^\m (z)$ and $Q_\n^\m (z)$ are Legendre functions \cite{as},
\ba
&&
z = \coth (\phi/\phi_1), \ \
\phi_1 \equiv a (2-\mu)=
\frac{4 a}{a^2+1}
, \lb{definz} 
\\&&
\mu \equiv 2 (a^2 - 1)/(a^2+1), \lb{definmu} 
\ea
and $\s$'s and $C$'s are integration constants.


Before going  further, let us emphasize the generic features of 
the constructed model, i.e., the one given 
by Eq. (\ref{eEMD}) with $\Lambda(\phi)$ and $\Xi(\phi)$
specified above.
The first thing to notice is that we actually did not use any specific high-D field
action to derive Eqs. (\ref{eCE1}), (\ref{eCE2}) - their derivation was based
solely on Eq. (\ref{eBC}).
Thus, what we have derived is the \textit{maximally general} 4D theory
of type (\ref{eEMD}) 
which can have in principle the brane-like solutions.
According to \scopr, the corresponding sector of the genuine EFMT must be a subset
of this theory: the principle states that the necessary condition to be EFMT is to belong to the p-brane class.

One may wonder, however, whether this brane class is unique: what if there 
exists other class whose $\phi(e^A)$ is different from Eq. (\ref{eBC}) thus
leading to the equations different from Eqs. (\ref{eCE1}), (\ref{eCE2})
and, therefore, resulting in different coupling functions?
The answer is: from what we said after Eq. (\ref{eBC}) it
follows that the class (\ref{eBC}) \textit{is} universal 
(in fact, we have checked that all known p-brane solutions related to M-theory do belong to this class), 
but even if this happens to be
not true - ambiguity will not increase but, quite the contrary, 
it will decrease further more.
The reason is that if an additional class does exist 
then the EFMT should lie on the intersection
of the new class and the old one (\ref{eBC}), 
therefore, the number of restricting 
equations would increase.
In principle, the only way to correct the theory 
(apart from introducing new fields into Eq. (\ref{eEMD})
which is a separate task) 
would be to generalize it by finding the  
class characteristic function $A (\phi)$ more general than Eq. (\ref{eBC}).
However, for that to happen new types of p-branes 
should be discovered because
the known ones do not suggest any necessity for such generalization.

The second point is despite the brane we used 
was derived from a perturbative low-energy theory (which
was used rather as a seed for the reconstruction formalism), 
it does not mean that the relation (\ref{eBC}) is restricted to this 
brane or to the perturbative theory as such.
In particular, it holds also for extremal branes which are
known to represent non-perturbative aspects 
(BPS saturation is protected from quantum corrections \cite{Witten:1978mh}).
In other words, full M-theory itself should also belong 
to this class to have brane solutions, therefore,  our EFMT 
provides a certain piece of information about the strong coupling regime.

\section{Eigenproblem in coupling space}\label{CHs}

In previous section we have determined the general functional form of couplings
required by \scopr.
The remaining problem is the value of the constant parameter $a$.
In principle, the extended SUGRA's propose the concrete values of this parameter:
$a =0$, $1$, $1/\sqrt{3}$, $\sqrt{3}$ \cite{Duff:1996qp}.

There exists, however, an elegant and all-sufficient way of determining a value
of $a$ without   involving  any external arguments, such as supersymmetry
(which is especially helpful if 
M theory
has other physically relevant limits not necessary obeying supersymmetry).
In fact, the idea logically follows from what we were doing before.
We will show that considering the \textit{eigen}value problem for Eq. (\ref{eCE2}) inevitably  
leads to the unique discrete (moreover, finite) spectrum of $a$'s.

Indeed, as long as the self-coupling $\Lambda$ is given by a solution of a second-order linear homogeneous
differential equation, the space of its values (functional, in general case) can be endowed with 
an orthogonal basis, according to the Sturm-Liouville theory of self-adjoint differential operators.
Therefore, the coupling space $\{\Lambda[\phi(x)]\}$ is  Hilbert,
much alike in ordinary quantum mechanics.

There are, however, important differences between this space and that of quantum mechanical
state vectors:

(i) wave functions are defined on spacetime whereas
couplings are defined on the field space, therefore, some habitual physical concepts,
such as energy, ground state, state vector collapse, measurement, etc., may not be directly extrapolated, 

(ii) wave functions are additionally normalized as to obey the probability interpretation
whereas for coupling functions we usually do not need that,

(iii) in order to have a physically admissible field-theoretical action one has to impose the reality
of the couplings themselves, i.e., when doing a ``coupling-space quantization'' 
(speaking more precisely, determining the allowed values of the coupling functions' parameters)
one should restrict oneself with a ``physical'' subspace of the Hilbert coupling space,
\be\lb{realcond}
\Lambda (\phi) \subset \text{``physical''} \ \ \text{for all} \ \ \phi \in \Re\text{e};
\ \ a^2 > 0,
\ee
where ``physical'' means  a set of at least real, single-valued and regular in a finite domain functions of $\phi$.
The value $a = 0$ has been \textit{ab initio} excluded 
because 
in the original p-brane ansatz the
scalar would become a constant and thus decouple (in 4D it corresponds to the \reno case)
so that Eq. (\ref{eBC}) is inapplicable.
Also, notice that Eqs. (\ref{eCE1}), (\ref{eCE2})
are invariant 
under the replacement
$a \to 1/a$ 
up to the electric-magnetic duality transformation, therefore, 
in the $a$ parameter space
it suffices
to consider 
either the domain $a^2 \leq 1$ 
with the removed point $a=0$
or 
the  $a^2 \geq 1$ one.

By transformation (see Eqs. (\ref{definz}), (\ref{definmu}) for notations)
\be
\Lambda(\phi) \to \Lambda(z)=w (z)/(1-z^2),
\ee
the differential equation for $\Lambda$ reduces to the Legendre one,
\be
(1-z^2)w''(z) - 2 z w'(z) + [6-\mu^2/(1-z^2)] w(z)=0,
\ee
such that
\be
w(z)= C_1 P_2^\mu (z)
+
C_2 Q_2^\mu (z),
\ee
so the general solution (\ref{eV-G}) could be eventually reproduced.
If we assume the physical range domain of $\phi$ to be a real axis
then via Eq. (\ref{definz}) the range of $z$ is $(-\infty,-1] \cup [1,\infty) $
where the origin in $\phi$-space corresponds to infinity in $z$-space
times a sign of $a$,
whereas the infinities in $\phi$-space correspond to $z=\pm\, \text{sgn}\, a$, respectively.

What about boundary conditions for the last differential equation?
Rigorously speaking, in general we do not have any.
However, 
the general solution for $w(z)$ has a cut
along the real axis for large values of $|z|$ (which corresponds to  small  $|\phi|$'s), 
therefore, 
when keeping  $C_i$ arbitrary
the  condition (\ref{realcond}) can be satisfied 
by their proper redefinition only if 
$\mu$ takes integer values (see Appendix \ref{App-mu} for details):
\be\lb{spectr2}
\mu = n_\mu = 0, \, \pm 1, \, \pm 2,\, \pm 3, ... \, ,
\ee 
though in principle negative values of $n_\mu$ can be discarded because they correspond
to $1/a$ spectrum, according to the property:
\be\lb{muadual}
\mu \to - \mu \ 
\Leftrightarrow \  
a\to 1/a,
\ee
see also the comment after Eq. (\ref{realcond}).
In its turn, the discretization of the $\mu$ spectrum  
implies that the Legendre functions in $w(z)$
become the associated Legendre polynomials,
such that, e.g., at $\mu \geq 0$ one can use the formulae:
\ba
&&
P_2^\mu(z) \to \frac{3}{2} (z^2-1)^{\mu/2} \frac{d^\mu}{d z^\mu} ( z^2-1/3),
\nn\\&&
Q_2^\mu(z) \to \frac{3}{2} (z^2-1)^{\mu/2} \frac{d^\mu}{d z^\mu} 
\left[
\frac{z^2-1/3}{2} 
\ln{\left(\frac{z+1}{z-1}\right)}
-  z
\right]\!,
\nn
\ea
which turn to hyperbolic trigonometric functions upon transforming 
back to $\phi$ by virtue of Eq. (\ref{definz}).

\begin{table}[t]
    \begin{tabular}{cccc}
      \hline \hline
      $~~n_\mu$ ~ & $a^2$ &~~~  $n_\mu$~ & $a^2$\\
      \hline
      $0$ & $1$ &  & \\
      $1$ & $3$ &  $-1$ & $1/3$ \\
      $2$ & $\infty$ &  $-2$ & $0$ \\
      $3$ & $-5$ &  $-3$ & $-1/5$ \\
      $4$ & $-3$ &  $-4$ & $-1/3$ \\
      $5$ & $-7/3$ &  $-5$ & $-3/7$ \\
      $\vdots$ & $\vdots$ &  $\vdots$ & $\vdots$ \\
      $\infty$ & $-1$ &  $- \infty$ & $-1$ \\
      \hline \hline
    \end{tabular}
    \caption{Correspondence between the spectra of $\mu$ and $a^2$.}
    \label{tab:mua}
\end{table}

The relation between $a^2$ and $\mu$ is given by Eq. (\ref{definmu})
so we can draw the Table \ref{tab:mua}
from which it is clear that under the assumption of reality of $a$, Eq. (\ref{realcond}),
 the only admissible values of $a^2$ are $0$, $1$ and $3$ or $1/3$
(though, in accordance with Eq. (\ref{muadual}), the first and last values are dual
to $\infty$ and $3$, respectively).
Thus, by Eq. (\ref{spectr2}) we have derived the relation 
noticed by Hull and Townsend 
when dealing with extreme black holes in 4D string compactifications \cite{Hull:1994ys}.
Thus, we have computed the ``magic numbers'' of extended supergravity theories  
from the model acting in the 
product coupling space 
$\{ \Lambda[\phi(x)] \} \otimes \{ \Xi[\phi(x)] \}$.

The corresponding eigenfunctions can be easily found by substituting eigenvalues into
the general solution  (\ref{eV-G}).
As was already mentioned, for the 
allowed values of $a$ the Legendre functions become just polynomials 
that greatly simplifies treatment. 
For instance, the case $a=1$ will be performed in the next section,
although in string theory it is the $a=\sqrt{3}$ one which is usually regarded as describing
an ``elementary'' black hole, in
a sense that other states, $1$, $1/\sqrt 3$, $0$, can be sequentially created 
by binding of, respectively, two, three and four ``elementary'' black holes \cite{Duff:1996qp,DL:1999}.

Thus, in general case  the scalar sector of EFMT is a superposition of
six Lagrangians  corresponding to 
the above-mentioned eigenvalues of $a$. 
The question which of them corresponds to ``ground state'', ``first excited'', etc.,
depends on whether it is possible to uniquely introduce the notion of energy
in the coupling space.
If it can be defined as in Appendix \ref{App-ene} then one can show
that if one assumes that $\phi$
plays a role of ``time'' in the coupling space 
then the states with $a = \pm 1$ indeed have larger $k(\phi)$ for all $\phi$ than others,
and thus are less favorable ``energetically''.

\section{EFMT and fundamental solution}

Now it is time to explicitly obtain 
that 4D solution whose high-D analogues are p-branes like the one 
given by Eqs. (\ref{st-pbrme}), (\ref{st-pbrphi}).
Henceforth we will concentrate solely on the $a=1$ case
(solution for arbitrary $a$ is given in the Appendix \ref{App-gss}).
Also, we assume the vanishing magnetic charge, $P=0$, hence $\hat\Xi$
becomes simply $\Xi^{-1}$ up to the constant factor $2 Q^2$ which cancels out anyway
(then the electric charge also does not appear in the action, as expected).
Then Eqs. (\ref{eXi-G}), (\ref{eV-G}) after substituting $a=1$ and redefining
integration constants  yield 
\be
1/\Xi = 
\frac{1}{2}
\left(
\s_1 e^{\phi} + \s_2 e^{-\phi}
\right)
,
\label{eXi}
\ee
\vspace{-5mm}
\be
\Lambda = - 
2 \lambda (\cosh \phi + 2)
- 4 \chi (3 \sinh \phi - \phi (\cosh \phi + 2))
,
\label{eV}
\ee
where $\s_i$, $\chi$ and $\lambda$ are arbitrary integration constants,
$\lambda$ is known as the ``cosmological constant'' parameter.
In principle, 
if one chooses $Q$ to vanish instead of $P$ then  $1/\Xi$ above should
be replaced  with $\Xi$ as the EM duality suggests
(and one should also reverse the sign of $\phi$ in $\Xi$). 

Notice that the derived model (\ref{eXi}), (\ref{eV}) generalizes not only the tree-level
case (\ref{e-minstr}) \cite{Gibbons:1987ps,Garfinkle:1990qj}
but also the  heuristic beyond-tree-level Monni-Cadoni's S-dual model 
(invariant under $\phi \to - \phi$) \cite{Monni:1995vu}, for which:
\be                                              
1/\Xi_{(MC)}^{\varsigma } = - \cosh{a \phi}, \ \
\Lambda_{(MC)} = 0,
\ee
(they considered the magnetic case where $\varsigma=-1$).
Besides, the potential (\ref{eV}) generalizes also the 
monoscalar gauged $N=8$, $SO(8)$ SUGRA 
representing the massless $U(1)^4$ sector of 
the M-theory's $AdS_4 \times S^7$ vacuum \cite{deWit:1982ig,DL:1999}
where:
\be                                               
\Xi_{(S^7)} = 0, \ \
\Lambda_{(S^7)} = - 
2 \lambda (\cosh \phi + 2).
\ee
In this connection, one could ask whether the $\chi$-term in the potential (\ref{eV})
(odd under inversion of the scalar) is redundant. 
Clearly, it is not - later on
we will show that it is responsible for the compatibility of the scalar field
with black holes as it guarantees the existence of event horizon (and thus 
protects from a naked singularity \cite{Penrose:1964wq}) even in absence of the electromagnetic field.

Finally, utilizing  
Appendix \ref{App-gss}
we can derive the  fundamental 0-brane solution
of our EFMT (\ref{eEMD}), (\ref{eXi}), (\ref{eV}).
Indeed, it describes  a charged black hole with scalar hair and is 
given 
by
\ba
&&
\drm s^2 = - N^2  \drm t^2 +  \frac{\drm r^2}{N^{2}}    + 
R^2 (\drm \theta^2 + \sin^2\!\theta \, \drm \varphi^2), 
\label{eSolg}\\&&
e^\phi = H, \
{\cal A}_0 =
-\frac{Q}{2}
\left(
\frac{\s_1}{r}+
\frac{\s_2}{r+\k}
\right)
,
\label{eSol}
\ea
where 
$
N^2 = 1 - 2 \chi 
\left[
\kappa (r + \kappa/2) - R^2 
\ln H
\right]
+ 
\frac{Q^2}{\k}
\bigl(
\frac{\s_2}{r+\k}-
\frac{\s_1}{r}
\bigr)
-
\lambda R^2
,$
$
R=\sqrt{r (r + \kappa)}
$
and $H = 1+\kappa/r$ (it is given by Eq. (\ref{st-pbrphi}) 
at $\tilde{d}=1$).
The model also admits another solution which can be deduced
from above by the simultaneous  transformations 
$\{\phi \to - \phi, \chi \to - \chi,  \s_1 \leftrightarrow \s_2 \}$ 
because the Lagrangian has corresponding symmetry.
One can show that the solution retains an event horizon even if the gauge field is 
switched off which leads to important phenomenological consequences - the potential 
(\ref{eV}) 
has been  used for
the unified description of  
cosmological phenomena and black holes \cite{Zloshchastiev:2004ny}.
Besides, there it was explained why the cosmological constant (mentioned at
the top of paper) should be bound from above by some positive number
whose origin and physical meaning are related to black holes.

\section{Conclusion}

To conclude, above we have described how 
it may be possible to 
derive a class of effective theories representing
the low-energy 4D limit of M-theory.
Thereby, our method does not 
make explicit use of any \textit{predefined} 
compactification manifolds or other geometrical constructions
which presume one or another way of the decomposition of space dimensions into ``observable''
and ``unobservable'' ones.
The approach does not address the question why our everyday world
is essentially four-dimensional
(though, $D=4$ does look distinguishable from a certain point of view, see the bottom of Sec. \ref{sec:scp}, but the physical meaning of that fact is rather obscure so far).
Neither the approach refuses the compactification idea as such.
Instead, it
focuses on the \textit{generic} features of EFMT, i.e., those which are independent from
how the above-mentioned decomposition could be done.

Here we have demonstrated the use of the method (by deriving both a particular sector of EFMT
and its corresponding brane-like solution) for the 
simplest abelian case -
when bosonic sector contains  graviton, photon and scalar. 
The future directions of work would be to further generalize the skeleton model (\ref{eEMD})  
by engaging other fields 
(Yang-Mills, higher-rank tensors, spinors, etc.) 
which could be required for describing
physics at higher energies and shorter scales of length.

\begin{acknowledgments} 
I thank C. Chryssomalakos, H. Quevedo, A. G\"uijosa, P. Dawson, and J. Vergara for enlightening discussions and stimulating remarks
which greatly helped to improve the paper. 
\end{acknowledgments} 


\appendix
\section{Eigenvalues of ${\mu}$}\label{App-mu}

Let us demonstrate that imposing the 
condition (\ref{realcond}) immediately
leads
to discretization of the $\mu$ spectrum.
So, we first infer that $\Lambda$ tends to a finite real value when
approaching the $z=\infty$ (i.e., $\phi=0$) point.
Using the asymptotics of $w$ for large $z$, we obtain:
\be
\Lambda(z)
\propto
-
\text e^{i\pi\mu/2}
\frac{C_1 + i \pi C_2/2}{\Gamma(3-\mu)/3}
+O(1/z^{2})
,
\ee
such that the constants must be related by
\be\lb{c1c2A}
C_1 
+ i \pi C_2/2
= 
-
\text{e}^{-i\pi\mu/2} \Gamma(3 - \mu )\,{\Lambda_0}/3
 ,
\ee
where 
$\Lambda_0 \equiv \lim\limits_{z\to \infty} \Lambda (z) = \Lambda (\phi)|_{\phi=0}$ 
has physical meaning of the (effective) cosmological constant,
up to a numerical factor, and thus is real-valued. 

Further, assuming $\mu$ positive for definiteness (see Eq. (\ref{muadual})) and
using the asymptotic of $w$ for $z \to 1+0$, we obtain:
\be
\Lambda(z)
\propto
-\frac{   C_1 + 
        \pi C_2  \cot (\pi \mu )/2
      }
      {2^{1-\mu/2} \text{e}^{i \pi \mu/2 }
      \Gamma(1 - \mu )
      \left(  z-1 \right)^{1 + \mu/2}} 
,
\ee
such that we arrive at another relation for $C$'s:
\be\lb{c1c2B}
   C_1 + 
        \pi C_2 \cot (\pi \mu )/2
=      
- 2^{1-\mu/2} \text{e}^{i \pi \mu/2 }
      \Gamma(1 - \mu ) \Lambda_\infty
 ,
\ee
where $\Lambda_\infty \equiv \lim\limits_{z\to 1+0}  (z-1)^{1 + \mu/2} \Lambda (z)$ 
is another real-valued constant. 
Thus, with Eqs. (\ref{c1c2A}) and (\ref{c1c2B}) in hand one can always express the constants $C_1$ and $C_2$ in terms
of the real-valued ones $\Lambda_0$ and $\Lambda_\infty$.

Now, we are left with the imposing the 
condition (\ref{realcond})  in
two more limits, at $z\to -\infty$ and $z \to -1-0$.
The former yields obviously nothing new as it brings us back to Eq. (\ref{c1c2A}) whereas
considering the asymptotical behavior of $\Lambda (z)$ at $z \to -1-0$, we obtain:
\be
\Lambda(z)
\propto
-\frac{   
        \pi C_2  \csc (\pi \,\mu )
      }
      {2^{2-\mu/2} \text{e}^{i \pi \mu/2 }
      \Gamma(1 - \mu )
      \left(  -z-1 \right)^{1 + \mu/2}} 
,
\ee
which means that
\be\lb{c1c2C}
\frac{   
           \csc (\pi \,\mu ) C_2
      }
      { \text{e}^{i \pi \mu/2 }
      \Gamma(1 - \mu )
     } 
\ee
must be real-valued.
Expressing here $C_2$ in terms of $\Lambda_0$ and $\Lambda_\infty$
by virtue of Eqs. (\ref{c1c2A}) and (\ref{c1c2B}),
we conclude that 
\be
2^{\mu/2} 
( \mu-1  )
(  \mu-2  )
    {{\Lambda }_0} - 
  6\,\text{e}^{i \pi \mu }
  {{\Lambda }_{\infty}}
\ee
must be real-valued.
This is possible only if $\mu$ is integer.

In fact, one can easily prove a more general result:
let 
$L_\nu^\mu(z) \equiv  C_1 P_\nu^\mu (z) + C_2 Q_\nu^\mu (z)$ 
be a linear combination of the associated Legendre functions on the real axis, whereby $C$'s are arbitrary.
Then  
the function $L_\nu^\mu(z)$ is real, single-valued and regular
for all $z \in (-\infty,-1) \cup (1,\infty) $
iff $\mu$ and $\nu$ are integers.
The proof consists in expanding $L_\nu^\mu(z)$ in series in the
neighborhood of some conveniently chosen points and
imposing the reality of the leading-order term, next-to-leading, etc.

\section{General static solution}\label{App-gss}

Here we present the exact general spherically symmetric static solution of the Einstein-Maxwell-scalar
gravity (\ref{eEMD}) with the couplings $\Xi$ and $\Lambda$ given by Eqs. (\ref{eXi-G}) and (\ref{eV-G}),
assuming electric case, i.e., 
$\hat\Xi= \Xi^{-1}$ up to a constant factor.
In fact, according to the method of Ref. \cite{Zloshchastiev:2001da}, once a class is established 
and the class equation
is derived and solved, the task of finding expressions for metric, scalar $\phi$ and 
electrostatic potential ${\cal A}_0$ is straightforward. 

So, 
assuming 
that we are working in 
the gauge (\ref{eSolg}) we respectively find
\ba
&&
N^2 = 
(1- \m/2)^2 
\text{e}^{ \frac{\m \phi}{\phi_1}}
(1+N_{\s}/\kappa^2
- N_\Lambda R^2),
\\&&
R^2 =
\frac{\text{e}^{\phi/{a}}\,{\kappa }^2}
  {
\bigl( 1 - \text{e}^{\frac{2\,\phi }{{{\phi }_1}}}
\bigr)^2}, \ \
\text e^{\frac{2 \phi}{\phi_1}}=
1 + \frac{\kappa ( 1 - \mu/2 ) }{r}
, \
\\&&
{\cal A}_0=
-\frac{Q{\left( 1 - \mu /2 \right) }^2\,  
      }{2 r} 
\left[ 
\sigma_1 + 
\frac{\sigma_2}{1 + \frac{\kappa ( 1 - \mu /2)}{r} } 
\right],
\\&&
N_{\s} \equiv
\frac{2 Q^2 {{\sigma }_1}}
  {2 + \mu } 
\left( 1 - \text{e}^{\frac{2\,\phi }{\phi_1 }} \right)
+
\frac{2 Q^2\,{{\sigma }_2}}
  { 2 - \mu }
\left( 1 - \text{e}^{-\frac{2\,\phi }{\phi_1}} \right)
,
\nn
\ea
and $N_{\Lambda} \equiv
(z \, + \,\mu/2)
 \Lambda'(\phi)/\phi_1
-
\Lambda(\phi) /2
$
where it is assumed that one  substitutes Eq. (\ref{eV-G}) and its derivative
into this formula.
Not all of these solutions describe the black holes embedded in the asymptotically flat or dS/AdS
spacetime but those with $a^2=1$, $3$, $1/3$ definitely do \cite{Hull:1994ys}.

\section{Measure of energy in coupling space}\label{App-ene}

We have a differential equation for a coupling function which
takes values in the coupling space $\{ C_i [\phi(x)]\}$
lying ``on top'' of the space of  scalar field $\phi$.
Its general form is 
\be\lb{cphieq}
C''(\phi)+ w_1(\phi) C'(\phi) + w_2 (\phi) C(\phi) =0,
\ee
where functions $w_i (\phi)$ are given.
This equation can be treated as an evolution equation of some dynamical system
in some effective space
such that one can introduce a notion of energy.

It seems, however, that the way of choosing the appropriate system
is  not unique.
In particular, it depends upon whether variable $\phi$ is associated with
``space-like'' or ``time-like'' coordinate in that effective space.
In former case, one could regard Eq. (\ref{cphieq}) as the stationary \schrod equation
for an imaginary particle
where $\phi$ plays the role of the 1D coordinate $x$, so that one can determine
energy by comparison:
\be\lb{qmener}
2 m (E - U) 
\Leftrightarrow
k (\phi),
\ee
where
\be
k(\phi) \equiv 
w_2 (\phi) - \frac{1}{2}  w_1' (\phi)
- \frac{1}{4} w_1^2 (\phi).
\ee

However, as long as our $\phi$ has physical meaning of the inflaton from scalar-driven 
inflationary cosmology,
we know that it is correlated with the cosmological 
``arrow of time''.
Then Eq. (\ref{cphieq}) can be regarded as the  Euler-Lagrange equations 
for a classical-mechanical system where $\phi$ plays the role of time.
Let us make a substitution
\be
C (\phi) =X(\phi) e^{- \frac{1}{2} \int w_1 (\phi) d \phi}, \ \
\phi \to t,
\ee
then Eq. (\ref{cphieq}) becomes the one for the harmonic oscillator
with a time-varying spring constant:
\be
\ddot X(t) 
+ 
k(t)
X(t) =0, \lb{cphieq2}
\ee
from which one can sequentially derive the Lagrangian and Hamiltonian
functions.
Thus, in principle
one can use
\be
E = 
\frac{1}{2} X'(\phi)^2 + 
\frac{1}{2} k(\phi) X(\phi)^2,
\ee
as a measure of energy in the coupling space,
but as long as $C(\phi)$ and hence $X(\phi)$ are determined
up to a constant factor, sometimes it will be more convenient
to use the ``energy-per-square-distance'' value
\be
{\cal E}\equiv 
2 E/X^2 = 
\left[
\frac{d \ln C(\phi)}{d\phi} 
+
\frac{1}{2} w_1 (\phi)
\right]^2 + 
 k(\phi),
\ee
which is $C(\phi)$-scale-invariant.
It agrees with the estimation of energy based on the fact
that the effective oscillation frequency equals to square root
of the spring constant, thus minimum of energy corresponds to minimum
of $k(\phi)$.
By the way, notice that in the ``Schr\"odinger'' interpretation (\ref{qmener}) the
minimum of energy would correspond to maximum
of $k(\phi)$ unless $m<0$.

\section{Theories in coupling space and renormgroup approach}\label{App-rga}

The following is mainly a conjecture which lies somewhere outside the main line of paper and
thus can be omitted in first reading.
Yet, it could be a good starting point for those interested in studying possible
relationship between the renormgroup (RG) approach in quantum field theory
and  dynamical models acting on the space of field couplings.

Indeed, Eqs. (\ref{eCE1}), (\ref{eCE2}) are in fact describing the behavior of coupling
``constants'' 
(which have been promoted to locally defined values, of course) with respect to 
the fundamental scalar, $\phi$.
We 
can assume that $\phi$ also defines the energy scale,
i.e.:
\be
\phi \sim \ln{\mu/\mu_0},
\ee
where $\mu$ is a characteristic cut-off energy-momentum scale,
and
$\mu_0$ is a normalization constant.
To see more clearly a motivation,
notice first that Eqs. (\ref{eCE1}), (\ref{eCE2}) 
are homogeneous second-order ODE's of type (\ref{cphieq})
(actually, it is a common feature of the dynamical equations for coupling functions
because field-theoretical Lagrangians are obviously linear with respect to couplings as such).
It is well-known that from the two integration constants which appear in the solution
of such an ODE
only one characterizes the details of dynamical behavior whereas other constant 
just determines scale: 
$
C(\phi) = c_1 C_{(1)}(\phi) + c_2 C_{(2)}(\phi) = c_2 [\hat c_1 C_{(1)}(\phi) +  C_{(2)}(\phi)]
$.
Thus, all the essential
information contained in a second-order homogeneous ODE can
be encoded in an appropriately chosen first-order ODE. 
Indeed, by substitution 
\be
g (\phi)=d \ln{C (\phi)}/d \phi
\ee
the equation can be brought to the RG-like form:
\[
\mu d g/d \mu =
d g/d \phi  = \beta_g(\phi,g) \equiv -g^2-w_1(\phi) g - w_2(\phi).
\]
If one is unhappy with the explicit appearance of $\phi$ inside 
$\beta_g(\phi,g)$,
one can in principle solve the differential equation for $g(\phi)$,
invert the function,
and replace $\phi$ by $\phi(g)$ in the r.h.s. of the equation.
Doing that way, one will arrive at the equivalent ``$\mu$-free r.h.s.'' form
(as in the habitual RG formalism):
\be
\mu d g/d \mu = d g/d \phi  = \beta_g(g),
\ee
where 
$\beta_g(g) \equiv \beta_g(\phi(g),g)$, i.e., in general case the beta function
is not just a quadratic polynomial of $g$. 

Incidentally, one should not 
think that 
the conjectured relationship infers that
the RG flow equations can be derived solely from the (classical)
field equations.
Despite one can definitely say that about Eq. (\ref{e-g1a}), 
to obtain Eq. (\ref{joinedce})
one needs the 
additional 
(external) input - 
the class characteristic equation, Eq. (\ref{eBC4}), which
comes from the
brane physics.

\def\AnP{Ann. Phys.}
\def\APP{Acta Phys. Polon.}
\def\CJP{Czech. J. Phys.}
\def\CMPh{Commun. Math. Phys.}
\def\CQG {Class. Quantum Grav.}
\def\EPL  {Europhys. Lett.}
\def\IJMP  {Int. J. Mod. Phys.}
\def\JMP{J. Math. Phys.}
\def\JPh{J. Phys.}
\def\FP{Fortschr. Phys.}
\def\GRG {Gen. Relativ. Gravit.}
\def\GC {Gravit. Cosmol.}
\def\LMPh {Lett. Math. Phys.}
\def\MPL  {Mod. Phys. Lett.}
\def\Nat {Nature}
\def\NCim {Nuovo Cimento}
\def\NPh  {Nucl. Phys.}
\def\PhE  {Phys.Essays}
\def\PhL  {Phys. Lett.}
\def\PhR  {Phys. Rev.}
\def\PhRL {Phys. Rev. Lett.}
\def\PhRp {Phys. Rept.}
\def\RMP  {Rev. Mod. Phys.}
\def\TMF {Teor. Mat. Fiz.}
\def\prp {report}
\def\Prp {Report}

\def\jn#1#2#3#4#5{{#1}{#2} {\bf #3}, {#4} {(#5)}} 

\def\boo#1#2#3#4#5{{\it #1} ({#2}, {#3}, {#4}){#5}}



\end{document}